\documentclass[aps,amsmath,amssymb,twocolumn,showpacs,pra]{revtex4-1}
\usepackage{graphicx}

\begin{document}

\title{Why state of quantum system is fully defined by density matrix}

\author{Alexey N. E. Nashev}
\email {nenashev@isp.nsc.ru}
\affiliation{Rzhanov Institute of Semiconductor Physics, 630090 Novosibirsk, Russia}
\affiliation{Novosibirsk State University, 630090 Novosibirsk, Russia}

\date{\today}

\begin{abstract}
We show that probabilities of results of all possible measurements performing on a quantum system 
depend on the system's state only through its density matrix. Therefore all experimentally available 
information about the state contains in the density matrix. 
In this study, we do not postulate that measurements obey some given formalism 
(such as observables, positive-operator valued measures, etc.), and do not use Born rule. 
The process of measurement is considered in a fully operational manner---as 
an interaction of a measured system with some black-box apparatus. 
The key point of our approach is the proof that, for improper mixtures, the expected value 
of any measurement depends linearly on the reduced density function. 
Such a proof is achieved by considering appropriate thought experiments. 
We demonstrate that Born rule can be derived as a natural consequence of our results. 
\end{abstract}


\maketitle

\section{Introduction}

When density matrices were invented (by Landau~\cite{Landau1927} and von Neumann~\cite{Neumann1927} in 1927), only one 
kind of quantum-mechanical measurements was known---namely, measurements of \emph{observables}. 
The observables are characterized by Hermitian operators, and probabilities of their measurement's 
results obey the Born rule,
\begin{equation} \label{eq:born-rule}
  \langle x \rangle \equiv \sum_k x_k P_k = \langle\psi| \hat x |\psi\rangle ,
\end{equation}
where $\langle x \rangle$ is the expected value of the observable $x$, $\hat x$ is the corresponding 
Hermitian operator, $P_k$ is the probability of the result $x_k$, and $|\psi\rangle$ is the state vector 
of the measured system. The rule~(\ref{eq:born-rule}) implies that the system is in a \emph{pure state}. 
In the case of a \emph{mixed state}, the concept of density matrix provides the natural 
generalization of the rule~(\ref{eq:born-rule}):
\begin{equation} \label{eq:born-rule-rho}
  \langle x \rangle = \text{Tr} \left( \hat \rho \hat x \right) ,
\end{equation}
where $\hat\rho$ is the density operator (density matrix) of the given state, pure or mixed. 
The way leading from Eq.~(\ref{eq:born-rule}) to Eq.~(\ref{eq:born-rule-rho}) is well known~\cite{Dirac,Landau}, 
and can be found in numerous textbooks. Since Eq.~(\ref{eq:born-rule-rho}) is valid for any 
observable $x$, then one can come to the following conclusion.

$\bullet$~{\bf Statement~$\boldsymbol{\rho}$.} \emph{Probabilities of measurement results depend on 
the state of the measured system only through the system's density matrix~$\hat\rho$.}

But this way of reasoning is restricted by measurements of observables. What about measurements of 
\emph{general} type, not necessarily describing by Born rule? In this paper, we will show that 
Statement~$\rho$ will remain in force for any thinkable measurement of quantum system.

The aim of the present study is to prove Statement~$\rho$ for the broadest 
possible class of measurements, and without any referring to Born rule. 

A broad class of measurements described by so-called positive operator valued measures (POVMs) 
is known since 1960s~\cite{Ludwig1964,Jauch1967}. 
By virtue of Neumark's theorem~\cite{Neumark1940}, such POVM measurements can 
be implemented by measuring appropriate observables of a larger quantum system; therefore they also 
obey Statement~$\rho$. But it is not clear \emph{a priori}, whether any possible measurement protocol can be 
expressed in terms of POVMs or observables, and this is the reason for reconsidering Statement~$\rho$ 
in the present paper. Our study will provide the positive answer to the latter question---see Section~\ref{discussion}.

\begin{figure}
\includegraphics[width=\linewidth]{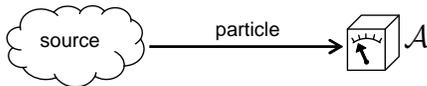}
\caption{Schematic representation of measurement by a ``black-box'' apparatus ${\cal A}$. 
The quantum system under measurement is symbolized as a particle, whose trajectory ends at the device ${\cal A}$.}
\label{fig:1}
\end{figure}

For the sake of generality, we will consider the process of measurement in a fully operational manner---as 
an interaction of a measured system with some black-box apparatus, see Fig.~\ref{fig:1}. 
For clarity, we will think of the measured system as of a moving particle that was emitted by some source 
and arrives at the measuring apparatus. We are not interested in any details of the apparatus construction 
and principle of operation, not even in the question which quantity is actually measured. 
Also, a fate of the particle after the measurement will not considered. We are concerned  
only with the measurement result, which can be read from the apparatus' scale. (Of course, it is assumed that 
by the time of the measurement the apparatus forgets about previously measured particles. Alternatively, 
one can suppose that for each measurement a new apparatus is taken.)
Such (or similar) operational point of view was used to give a description of effects and operations 
in quantum mechanics~\cite{Kraus,Holevo}, as well as to reconstruct quantum theory from simple 
axioms~\cite{Hardy2001,Hardy2013}.

It is important for our study to distinguish between \emph{proper} and \emph{improper} mixed states, 
in terminology introduced by d'Espagnat~\cite{dEspagnat1999}. A proper mixture arises when the quantum system 
is prepared in an unknown pure state, but probabilities of occurring of each pure state are known. 
For example, a fully non-polarized electron spin, described by the density matrix
\begin{equation} \label{eq:01}
  \hat\rho_{\text{non-polariz}} = 
  \left(  
  \begin{matrix}
  1/2 & 0\\
  0 & 1/2
  \end{matrix}
  \right) ,
\end{equation}
can appear (as a proper mixture) from taking spin-up and spin-down states with equal probabilities $1/2$:
\begin{equation} \label{eq:02}
  \hat\rho_{\text{non-polariz}} = \frac12
  \left(  
  \begin{matrix}
  1 & 0\\
  0 & 0
  \end{matrix}
  \right) 
  + \frac12
  \left(  
  \begin{matrix}
  0 & 0\\
  0 & 1
  \end{matrix}
  \right) .
\end{equation}
An improper mixture occurs when two quantum system are entangled, and the state of one of these systems 
\emph{alone} is considered. 
As an example of an improper mixture, one can view a pair of electrons in the spin-singlet state, 
\begin{equation} \label{eq:03}
  |S\rangle = 
  \frac{|\!\!\uparrow\rangle|\!\!\downarrow\rangle-|\!\!\downarrow\rangle|\!\!\uparrow\rangle}{\sqrt 2} ,
\end{equation}
and ask in which state the first electron is. According to the standard procedure of taking a partial trace, 
the (reduced) density matrix of the first electron's spin is
\begin{equation} \label{eq:04}
  \text{Tr}_{\text{2nd spin}} 
  \left(  
  \begin{matrix}
  0 & 0 & 0 & 0 \\
  0 & 1/2 & -1/2 & 0 \\
  0 & -1/2 & 1/2 & 0 \\
  0 & 0 & 0 & 0 
  \end{matrix}
  \right) 
  = \hat\rho_{\text{non-polariz}} ,
\end{equation}
i.~e. the first electron is non-polarized. 
Note that both kinds of mixtures ultimately originate from \emph{pure states}, 
therefore the probability rules for mixtures follow from the rules for pure states.

These trivial examples show that the same density matrix $\hat\rho_{\text{non-polariz}}$ may appear 
as a result of either a proper mixture, of an improper one. Moreover, different proper mixtures may 
result to the same density matrix: e.~g. the density matrix $\hat\rho_{\text{non-polariz}}$ may 
arise from mixing of spin-up and spin-down states according to Eq.~(\ref{eq:02}), as well as 
from mixing of spin-left and spin-right states 
$(|\!\!\uparrow\rangle|-|\!\!\downarrow\rangle)/\sqrt 2$ and $(|\!\!\uparrow\rangle|+|\!\!\downarrow\rangle)/\sqrt 2$:
\begin{equation} \label{eq:05}
  \hat\rho_{\text{non-polariz}} = \frac12
  \left(  
  \begin{matrix}
  1/2 & -1/2\\
  -1/2 & 1/2
  \end{matrix}
  \right) 
  + \frac12
  \left(  
  \begin{matrix}
  1/2 & 1/2\\
  1/2 & 1/2
  \end{matrix}
  \right) .
\end{equation}
This is also true for improper mixtures---for example, the singlet state $|S\rangle$, 
Eq.~(\ref{eq:03}), and the triplet state with spin projection 0,
\begin{equation} \label{eq:06}
  |T_0\rangle = 
  \frac{|\!\!\uparrow\rangle|\!\!\downarrow\rangle+|\!\!\downarrow\rangle|\!\!\uparrow\rangle}{\sqrt 2} ,
\end{equation}
of a pair of electrons 
result in the same density matrix $\hat\rho_{\text{non-polariz}}$ of one electron. 
Hence, in order to justify Statement~$\rho$, one should address the following three questions.

(Q1) Whether different \emph{improper} mixtures corresponding to the same density matrix are indistinguishable?

(Q2) Whether different \emph{proper} mixtures corresponding to the same density matrix are indistinguishable?

(Q3) Whether a proper mixture and an improper one are indistinguishable, if they correspond to the same 
density matrix?

(``Indistinguishability'' is meant here in the statistical sense, as equal probabilities of passing any 
possible experimental tests.)

These questions will be considered in details in the main part of the paper, 
Sections~\ref{improper1}--\ref{general}. 
But at first we will briefly outline our approach, using the spin states discussed above as an example. 

Let us ensure that two-electron states $|S\rangle$ and $|T_0\rangle$, Eqs.~(\ref{eq:03}) and~(\ref{eq:06}), cannot 
be distinguished from each other by any tests on the first electron. Indeed, the state $|S\rangle$ can be transformed into 
$|T_0\rangle$ by applying the unitary operator $\hat\sigma_z$ to the spin of \emph{the second} electron 
(for example, by action of local magnetic field). Hence, if it were be possible to distinguish 
between $|S\rangle$ and $|T_0\rangle$ by measuring \emph{the first} electron, 
one could use this for instantaneous communication, which 
is unphysical. This argument was elaborated by Zurek~\cite{Zurek2003,Zurek2005} providing therefore the 
positive answer to question Q1---see details in Section~\ref{improper1}. 

For answering questions Q2 and Q3, we developed an original approach 
presented in Sections~\ref{improper2} and~\ref{general}. 
Here we will illustrate our method by the example of two above-mentioned proper mixtures: 
the mixture $\mathfrak{m}_1$ of spin-up and spin-down states with equal probabilities, Eq.~(\ref{eq:02}), 
and the analogous mixture $\mathfrak{m}_2$ of spin-left and spin-right states, Eq.~(\ref{eq:05}). 
Let us consider some experimental test of the spin degree of freedom, and denote the 
probability of passing this test as $P$. It follows from the affirmative answer to question Q1, 
that $P$ is a function of the density matrix \emph{for improper mixtures and pure states}. 
We will prove in Section~\ref{improper2} that this function is \emph{linear}. 
It is convenient to represent the density matrix $\hat\rho$ of the electron spin through the vector of spin polarization~$\mathbf{p}$, 
defined by the equality
\begin{equation} \label{eq:06a}
  \hat\rho = \hat\rho_{\text{non-polariz}} + \frac12 \left( p_x\hat\sigma_x + p_y\hat\sigma_y + p_z\hat\sigma_z \right),
\end{equation}
$\hat\sigma_x$, $\hat\sigma_y$, $\hat\sigma_z$ being the Pauli matrices. So $P$ is a linear function of $\mathbf{p}$:
\begin{equation} \label{eq:07}
  P (\mathbf{p}) = \mathbf{a}\cdot\mathbf{p} + b ,
\end{equation}
where $\mathbf{a}$ and $b$ are some coefficients. So far Eq.~(\ref{eq:07}) can be applied only to improper mixtures 
and to pure states: for example, in the spin-singlet state $|S\rangle$ the polarization vector $\mathbf{p}$ is equal to zero, 
therefore $P = b$. In the spin-up state $|\!\!\uparrow\rangle$, $\mathbf{p}=\mathbf{e}_z$, and $P = a_z+b$; 
similarly, in the spin-down state $|\!\!\downarrow\rangle$, $\mathbf{p}=-\mathbf{e}_z$, and $P = -a_z+b$. 
In the mixture $\mathfrak{m}_1$, the electron spin is either up or down, with equal probabilities $1/2$; thus the 
probability of passing the test is
\begin{equation} \label{eq:08}
  P (\mathfrak{m}_1) = \frac12 P (\mathbf{e}_z) + \frac12 P (-\mathbf{e}_z) = b .
\end{equation}
In the same way,
\begin{equation} \label{eq:09}
  P (\mathfrak{m}_2) = \frac12 P (\mathbf{e}_x) + \frac12 P (-\mathbf{e}_x) = b .
\end{equation}
One can see that $P (\mathfrak{m}_1) = P (\mathfrak{m}_2) = P (|S\rangle)$ 
\emph{for any experimental test}. Hence, no test can distinguish between proper mixtures $\mathfrak{m}_1$ and $\mathfrak{m}_2$ 
and the entangled state (improper mixture) $|S\rangle$.

The key point of our way of reasoning is the proof that the probability depends on the density function linearly. 
Such a proof will be achieved by considering two novel thought experiments in Section~\ref{improper2}.

The rest of the paper is organized as follows.  
In Section~\ref{improper1} we will see how Zurek's concept of \emph{envariance} gives a proof of 
Statement~$\rho$ for improper mixtures, and thus gives the answer to question Q1. 
Section~\ref{preparation} introduces some tools which will be used below. 
Section~\ref{improper2} is the central part of the paper; using the thought experiments shown in Fig.~\ref{fig:3},
linearity of expected values as functions of the density matrix will be established in this section for improper mixtures. 
On the basis of the latter result, mixtures of general type will be considered in Section~\ref{general}, 
that will complete the proof of Statement~$\rho$. 
Then, in Section~\ref{discussion}, we will discuss the place of our results among other assertions of quantum theory of measurement. 
In particular, we will demonstrate that Born rule can be considered as a consequence of Statement~$\rho$. 
Concluding remarks will be gathered in Section~\ref{conclusions}.

In this work we assume that the \emph{non-measurement} part of quantum mechanics is standard---in particular, 
pure states are described by vectors in Hilbert spaces; 
a composite system's state space is the tensor product of state spaces of its parts; 
the evolution of state vectors between measurements is unitary.


\section{Proof of Statement~$\boldsymbol{\rho}$ for improper mixtures}\label{improper1}

A derivation of Statement~$\rho$ \emph{for the special case of improper mixtures} can be found in 
Zurek's works on \emph{envariance}~\cite{Zurek2003,Zurek2005}. 
Zurek considered a quantum system ${\cal S}$ entangled with some environment ${\cal E}$, so that 
the combined system ${\cal SE}$ is in a pure state $\psi_{\cal SE}$. Such a joint state can be 
represented in the form of Schmidt decomposition,
\begin{equation} \label{eq:schmidt}
  |\psi_{\cal SE}\rangle = \sum_k \alpha_k |s_k\rangle |\varepsilon_k\rangle \, ,
\end{equation}
where unit vectors $|s_k\rangle$ are mutually orthogonal and belong to the system ${\cal S}$, 
as well as unit vectors $|\varepsilon_k\rangle$ are mutually orthogonal and belong to ${\cal E}$. 
Then, the following statement was formulated and proved in Ref.~\onlinecite{Zurek2005}:

``\emph{Theorem 1.} For an entangled global state of the system and the
environment all measurable properties of ${\cal S}$---including probabilities
of various outcomes---cannot depend on the phases of Schmidt coefficients:
The state of ${\cal S}$ has to be completely determined by the set of pairs
$\{|\alpha_k|,~|s_k\rangle\}$.''

The idea of the proof can be represented as follows. Any state $|\widetilde\psi_{\cal SE}\rangle$ having the form 
\begin{equation} \label{eq:schmidt-phase}
  |\widetilde\psi_{\cal SE}\rangle = \sum_k \exp(i\phi_k) \, \alpha_k |s_k\rangle |\varepsilon_k\rangle  \, ,
\end{equation}
where $\exp(i\phi_k)$ are arbitrary phase factors, can be converted into the state $|\psi_{\cal SE}\rangle$ 
by application of a unitary operator
\begin{equation} \label{eq:schmidt-env}
  \hat U_{\cal E} = \sum_k \exp(-i\phi_k) \, |\varepsilon_k\rangle \langle\varepsilon_k|  
\end{equation}
to the environment ${\cal E}$. Since the system ${\cal S}$ ``does not know'' whether the environment 
was subjected to action of $\hat U_{\cal E}$ or was not, this action cannot alter any properties of~${\cal S}$. 
Therefore the state of the system ${\cal S}$ must be one and the same for state vectors 
$|\psi_{\cal SE}\rangle$ [Eq.~(\ref{eq:schmidt})] 
and $|\widetilde\psi_{\cal SE}\rangle$ [Eq.~(\ref{eq:schmidt-phase})] of the combined system ${\cal SE}$.

After a proof of Theorem~1, Zurek~\cite{Zurek2005} placed the following

``\emph{Remark.} The information content of the list
$\{|\alpha_k|,~|s_k\rangle\}$ that describes the state of ${\cal S}$
is the same as the information content of the reduced density matrix.''

Thus, the proof of Theorem~1 also proves Statement~$\rho$ in the particular case of improper mixtures, 
i.~e. when a combined system ${\cal SE}$ is in a pure state (${\cal S}$ is the system under measurement, 
and ${\cal E}$ is some other quantum system). As a consequence, 
if a state of a quantum system ${\cal S}$ is an improper mixture, then the expected value ${\mathbb E}$ 
of a result of measurement on ${\cal S}$ by some apparatus ${\cal A}$ is a function 
of the system's reduced density matrix $\hat\rho$. Let us denote this function as $F$:
\begin{equation} \label{eq:F-def}
  {\mathbb E} = F(\hat\rho)  \, .
\end{equation}
The function $F(\hat\rho)$ depends on choice of measuring apparatus. 

One can see that the derivation of Statement~$\rho$ for improper mixtures is based on 
three assumptions listed below.

$\bullet$~{\bf Assumption~$\boldsymbol{\psi}$.} \emph{When a~measured quantum system is in a~pure state, 
its state vector~$|\psi\rangle$ contains all information about probabilities 
of~measurement outcomes.}

(Assumption~$\psi$ just asserts that there are no hidden parameters beyond the state vector.)

$\bullet$~{\bf Assumption~$\boldsymbol{U}$.} \emph{Any unitary operation in a state space of 
a quantum system can be implemented by an appropriate physical device acting to this system.}

$\bullet$~{\bf Assumption~$\boldsymbol{I}$.} \emph{Information transfer without physical interaction 
is impossible.}

Note that Statement~$\rho$, and thus Eq.~(\ref{eq:F-def}), is also valid for pure states 
by virtue of Assumption~$\psi$.


\section{Preparation of one system by measurement of another system} \label{preparation}

In this Section, we will introduce and discuss three auxiliary entities: 
(i) a quantum system with a two-dimensional state space (e.~g. a spin-1/2 particle), which we will call a \emph{qubit}; 
(ii) a measuring device $\mu$ (a ``meter'') that performs a projective measurement on a qubit in some 
orthonormal basis $\left\{ |0\rangle, |1\rangle \! \right\}$ (e.~g. a Stern--Gerlach apparatus for measuring a spin projection); 
(iii) a source $Q$ that emits on demand a pair of entangled qubits prepared in the Bell state
\begin{equation} \label{eq:phi}
  |\Phi\rangle = \left( |0\rangle |0\rangle + |1\rangle |1\rangle \right) / \sqrt 2 .
\end{equation}

\begin{figure}
\includegraphics[width=\linewidth]{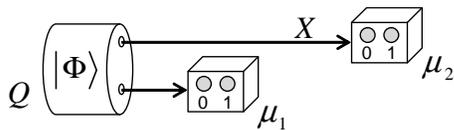}
\caption{Preparation of one system by measurement of another system. 
The source $Q$ emits two entangled particles (qubits) in the Bell state $|\Phi\rangle$, Eq.~(\ref{eq:phi}). 
Then the particles are subjected to measurements in the basis $\{ |0\rangle, |1\rangle \! \}$ 
by ``meters'' $\mu_1$ and $\mu_2$. 
Near the point $X$, the upper qubit appears in a pure state $|0\rangle$ or $|1\rangle$, depending on 
the result of the lower qubit's measurement. Arrows denote trajectories of the particles.}
\label{fig:2}
\end{figure}

Let us combine these things as shown in Fig.~\ref{fig:2}. Two qubits are prepared in the joint state 
$|\Phi\rangle$, then they are measured successively by ``meters'' $\mu_1$ and $\mu_2$. 
Looking at Eq.~(\ref{eq:phi}), one can see that the two measurements always give equal results. 
Therefore, when the upper qubit reaches the point $X$, one can \emph{predict} the result of its future 
measurement---namely, it must be equal to the result obtained by ``meter'' $\mu_1$. It is commonly believed 
that such a possibility of prediction means that the qubit appears to be in a definite, pure state $|x\rangle$, 
where $x$ is equal to the output of $\mu_1$. One can conclude hence that measurement of the lower 
qubit in the setup shown in Fig.~\ref{fig:2} \emph{prepares} the upper one in a pure state 
$|0\rangle$ or $|1\rangle$. 

Note that the following assumption is demanded for this way of reasoning:

$\bullet$~{\bf Assumption~$\boldsymbol{P}$.} \emph{If the probability of finding the qubit (by a projective 
measurement) in the state $|0\rangle$ (or~$|1\rangle$) is $1$, 
then the qubit is actually in the state $|0\rangle$ (or~$|1\rangle$).}


\section{Two thought experiments}\label{improper2}

In this Section, we will see that the function $F(\hat\rho)$ introduced in Section~\ref{improper1} is 
\emph{linear}. More precisely, we will see that for any two density matrices $\hat\rho_0$, $\hat\rho_1$ and any 
real number $\lambda\in[0,1]$ 
\begin{equation} \label{eq:main}
  F\left[ (1-\lambda)\hat\rho_0 + \lambda\hat\rho_1 \right] 
  = (1-\lambda) F(\hat\rho_0) + \lambda F(\hat\rho_1) \, .
\end{equation}
Equation~(\ref{eq:main}) is the central result of the present paper. It provides an opportunity to generalize 
Statement~$\rho$ (that is yet justified only for improper mixtures) to arbitrary mixtures, as we will see 
in Section~\ref{general}.

\begin{figure}
\includegraphics[width=\linewidth]{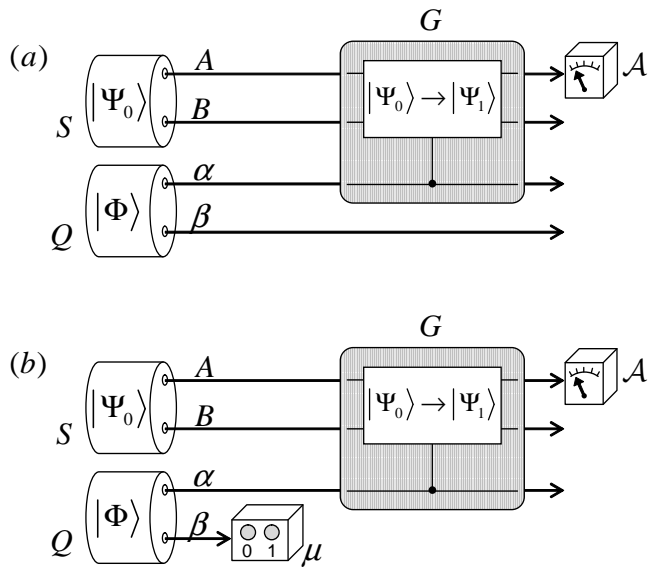}
\caption{Two thought experiments used in a proof of Eq.~(\ref{eq:main}). 
Experiment $(a)$: source $S$ emits a pair of particles $A,B$ in the joint state $|\Psi_0\rangle$ and, 
simultaneously, source $Q$ emits a pair of qubits $\alpha,\beta$ in the 
Bell state $|\Phi\rangle$, Eq.~(\ref{eq:phi}). 
Three particles $A,B,\alpha$ pass through gate $G$, that operates according to Eqs.~(\ref{eq:g0}),~(\ref{eq:g1}), 
and particle $A$ is measured by apparatus ${\cal A}$. 
Experiment $(b)$: the same as the previous experiment, but qubit $\beta$ is measured by ``meter'' $\mu$ 
in the basis $\{ |0\rangle, |1\rangle \! \}$.}
\label{fig:3}
\end{figure}

Let~$\hat\rho_0$ and~$\hat\rho_1$ be two density matrices of~some particle~$A$. 
One can always choose such a particle~$B$ and such two \emph{pure} states~$|\Psi_0\rangle$ and~$|\Psi_1\rangle$ 
of the combined system~$A+B$, that the~reduced density matrix of~$A$ 
is equal to~$\hat\rho_0$ for the state~$|\Psi_0\rangle$, and to~$\hat\rho_1$ for the state~$|\Psi_1\rangle$. 
Then, consider a~thought experiment shown in Fig.~\ref{fig:3}a. The source~$S$ emits a~pair of 
particles~$A$ and~$B$ prepared in the state~$|\Psi_0\rangle$. Simultaneously, another source~$Q$ 
emits a~pair of entangled qubits $\alpha$ and~$\beta$ in the Bell state $|\Phi\rangle$, Eq.~(\ref{eq:phi}). 
Then three particles~$A$, $B$ and~$\alpha$ go through a ``quantum gate''~$G$ that performs the~following 
``controlled transformation'':
\begin{align} 
  \label{eq:g0}
  |\Psi_0\rangle |0\rangle  &\stackrel{G}{\to}  |\Psi_0\rangle |0\rangle , \\
  \label{eq:g1}
  |\Psi_0\rangle |1\rangle  &\stackrel{G}{\to}  |\Psi_1\rangle |1\rangle .
\end{align}
(Such transformation can be done by an unitary operation, and thus can be physically realized according to 
Assumption~$U$.) 
Finally, the particle~$A$ is measured by the~same apparatus~${\cal A}$ that was considered above. 
We are interested in the expected value ${\mathbb E}$ of the measurement result.

Just after leaving the sources, the~joint state of four particles~$A,B,\alpha,\beta$ is 
\[
  |\Psi_0\rangle |\Phi\rangle  \equiv  
  \frac{ |\Psi_0\rangle |0\rangle |0\rangle + |\Psi_0\rangle |1\rangle |1\rangle }{\sqrt 2} \, .
\]
The gate~$G$ transforms it into the following state:
\[
  \frac{ |\Psi_0\rangle |0\rangle |0\rangle + |\Psi_1\rangle |1\rangle |1\rangle }{\sqrt 2} \, .
\]
For the latter state, the~reduced density matrix of the particle~$A$ is $(\hat\rho_0 + \hat\rho_1)/2$.
Therefore, according to Eq.~(\ref{eq:F-def}), 
\begin{equation} \label{eq:p-m-flash-a}
  {\mathbb E} = F \left( \frac{\hat\rho_0 + \hat\rho_1}{2} \right) .
\end{equation}

Now we will consider a~modification of this experiment shown in Fig.~\ref{fig:3}b. The~only difference between 
Figs.~\ref{fig:3}a and~\ref{fig:3}b is that, in the~latter experiment, the qubit~$\beta$ is measured in the 
basis~$\{|0\rangle,|1\rangle\}$ (by the ``meter''~$\mu$) before the~rest three particles reach the gate~$G$. 

In both experiments, the~trajectory of particle~$\beta$ is 
spatially separated from (and is not interacting~to) the~rest of the~setup, therefore 
due to Assumption~$I$ no information 
about this particle can reach the~measuring device~${\cal A}$. Consequently the expected value ${\mathbb E}$ 
of the measurement result is the~same for both experiments. 

On the other hand, the value of ${\mathbb E}$ in the second experiment can be expressed via 
the law of total expectation. Let ${\mathbb E}_0$ (${\mathbb E}_1$) be the conditional expectation of apparatus ${\cal A}$'s 
result given that meter $\mu$ provides the result 0 (1). Then, 
\begin{equation} \label{eq:total-probability}
  {\mathbb E} = (1-a) \, {\mathbb E}_0 + a \, {\mathbb E}_1 ,
\end{equation}
where $a$ is the probability of obtaining 1 by meter $\mu$, and $1-a$ is the probability 
of obtaining 0. Note that the quantity $a$ does not depend on states~$|\Psi_0\rangle$ and~$|\Psi_1\rangle$. 

Now we will show how the quantities ${\mathbb E}_0$ and ${\mathbb E}_1$ can be evaluated through the function $F$. 
Let the meter $\mu$ gave the answer 0. Then, according to our consideration in Section~\ref{preparation}, 
qubit $\alpha$ appears in the state $|0\rangle$ just after the measurement of qubit $\beta$. 
Due to Eq.~(\ref{eq:g0}), in this case the particles~$A$ and~$B$ will remain to be in the 
state~$|\Psi_0\rangle$ after passing through the gate~$G$. The~partial density matrix of the particle~$A$ 
before its measurement will be equal to~$\hat\rho_0$. Thus, according to Eq.~(\ref{eq:F-def}),
\begin{equation} \label{eq:p-m-flash-mu0}
  {\mathbb E}_0 = F(\hat\rho_0) .
\end{equation}
Similarly, if the result of measurement of qubit~$\beta$ was~1, then the qubit~$\alpha$ will be in the 
state~$|1\rangle$ after this measurement. The gate~$G$ will change the~state of 
particles~$A$ and~$B$ from~$|\Psi_0\rangle$ to~$|\Psi_1\rangle$, according to Eq.~(\ref{eq:g1}), and 
the~partial density matrix of the particle~$A$ before its measurement will be equal to~$\hat\rho_1$. Hence,
\begin{equation} \label{eq:p-m-flash-mu1}
  {\mathbb E}_1 = F(\hat\rho_1) .
\end{equation}

Taking equations (\ref{eq:p-m-flash-a}), (\ref{eq:p-m-flash-mu0}), (\ref{eq:p-m-flash-mu1}) into account, 
one can rewrite Eq.~(\ref{eq:total-probability}) in terms of the function $F(\hat\rho)$:
\begin{equation} \label{eq:f-p1}
  F \left( \frac{\hat\rho_0 + \hat\rho_1}{2} \right) = 
  (1-a) \, F(\hat\rho_0) + a \, F(\hat\rho_1) .
\end{equation}
Since this equation is valid for any two density matrices $\hat\rho_0$, $\hat\rho_1$, swapping 
$\hat\rho_0$ and $\hat\rho_1$ also provides a correct equation:
\begin{equation} \label{eq:f-p2}
  F \left( \frac{\hat\rho_1 + \hat\rho_0}{2} \right) = 
  (1-a) \, F(\hat\rho_1) + a \, F(\hat\rho_0) .
\end{equation}
Summing up Eqs.~(\ref{eq:f-p1}) and~(\ref{eq:f-p2}), one can make sure that  
\begin{equation} \label{eq:f-one-half}
  F \left( \frac{\hat\rho_0 + \hat\rho_1}{2} \right) = 
  \frac{ F(\hat\rho_0) + F(\hat\rho_1) }{2} 
\end{equation}
for any density matrices $\hat\rho_0$ and $\hat\rho_1$. 
This justifies Eq.~(\ref{eq:main}) for the particular case of $\lambda=1/2$. 

Iterating Eq.~(\ref{eq:f-one-half}), one can prove Eq.~(\ref{eq:main}) for 
infinitely many values of $\lambda$ within the range $[0,1]$. 
For example, let us see that Eq.~(\ref{eq:main}) is correct at $\lambda=1/4$:
\begin{multline*}
  F\left( \frac34\, \hat\rho_0 + \frac14\, \hat\rho_1 \right) 
  = F\left( \frac{ \hat\rho_0 + \frac{ \hat\rho_0 + \hat\rho_1 }{2} }{2} \right) \\
  = \frac{ F(\hat\rho_0) + F\left(\frac{ \hat\rho_0 + \hat\rho_1 }{2}\right) }{2} 
  = \frac{ F(\hat\rho_0) + \frac{ F(\hat\rho_0) + F(\hat\rho_1) }{2} }{2} \\
  = \frac34\, F(\hat\rho_0) + \frac14\, F(\hat\rho_1) \, .
\end{multline*}
This method allows to prove Eq.~(\ref{eq:main}) 
for any $\lambda=p/2^q$, where $q=1,2,3,\ldots,$ and $p=0,1,2,\ldots,2^q$. 
The set $\Lambda$ of all such values of $\lambda$ is a dense subset of the range $[0,1]$. 
Therefore, if the function $F$ is continuous, then the (given above) proof of Eq.~(\ref{eq:main}) 
for all $\lambda\in\Lambda$ automatically justifies Eq.~(\ref{eq:main}) for all $\lambda\in[0,1]$.

Hence, we have proven Eq.~(\ref{eq:main}), assuming that $F(\hat\rho)$ is a continuous function. 
However, one can go further and cancel the assumption of continuity of this function, 
considering slightly modified version of the thought experiments. 
This issue is discussed in Appendix~\ref{app}.

In the next Section we will see how Eq.~(\ref{eq:main}) helps to expand the range of applicability 
of Eq.~(\ref{eq:F-def}) to the case of general mixtures, that in its turn allows to justify Statement~$\rho$.


\section{General mixtures}\label{general}

A proper mixture takes place when a system ${\cal S}$ is in some quantum state, but we do not know which. 
Our ignorance of the actual state can be represented by a probability distribution over the different states. 
Generally, the actual (unknown) state may be not only a pure state of the system ${\cal S}$ alone, but 
a pure state of a larger system ${\cal SE}$, when ${\cal E}$ is some environment (so that the state of 
the system ${\cal S}$ is an improper mixture). We suppose that these options cover all possible cases 
of quantum states. In other words, we accept the following assumption.

$\bullet$~{\bf Assumption~$\boldsymbol{M}$.} \emph{Any state of a quantum system is either a pure state, 
or an improper mixture, or a mixture of pure states and/or of improper mixtures taken with some probabilities.}

Let us generalize Eq.~(\ref{eq:F-def}) to the case of states described in Assumption~$M$. 
Consider a probabilistic mixture of $N$ pure states of a combined system ${\cal SE}$, where ${\cal S}$ 
is the system under study. 
(Some of these states, or even all, may be pure states of the system ${\cal S}$ alone.) 
Each ($k$th) state is characterized by its reduced density matrix $\hat\rho_k$ in the state space of~${\cal S}$, 
and enters into the mixture with some probability $p_k$. The sum of probabilities is unity,
\begin{equation} \label{eq:sum-p-is-1}
  p_1 + \ldots + p_N =1 \, .
\end{equation}
By definition, the density matrix $\hat\rho$ of the whole mixture is equal to 
\begin{equation} \label{eq:rho-general}
   \hat\rho = \sum_{k=1}^N p_k \hat\rho_k \, .
\end{equation}
Let system ${\cal S}$ be measured by apparatus ${\cal A}$.  
Since each $k$th state is either an improper mixture or a pure state, one can apply the results of 
Section~\ref{improper1} and find that, for this state, the expected value ${\mathbb E}_k$ of the measurement result is 
\begin{equation} \label{eq:expected-kth}
  {\mathbb E}_k = F(\hat\rho_k) .
\end{equation}
The expected value ${\mathbb E}$ for the whole mixture can be found via the law of total expectation:
\begin{equation} \label{eq:total-probability-N}
  {\mathbb E} =  p_1 {\mathbb E}_1 + \ldots + p_N {\mathbb E}_N .
\end{equation}
Therefore, taking Eq.~(\ref{eq:expected-kth}) into account,
\begin{equation} \label{eq:total-probability-N2}
  {\mathbb E} = p_1 F(\hat\rho_1) + \ldots + p_N F(\hat\rho_N) \, .
\end{equation}
It is easy to check by induction, starting from Eq.~(\ref{eq:main}), that 
\begin{equation} \label{eq:main-N}
  p_1 F(\hat\rho_1) + \ldots + p_N F(\hat\rho_N) 
  = F\left( p_1 \hat\rho_1 + \ldots + p_N \hat\rho_N \right) \, .
\end{equation}
For example, when $N=3$
\begin{multline*} 
  p_1 F(\hat\rho_1) + p_2 F(\hat\rho_2) + p_3 F(\hat\rho_3) \\
  = p_1 F(\hat\rho_1) + (p_2+p_3) \left[ \frac{p_2}{p_2+p_3} F(\hat\rho_2) + \frac{p_3}{p_2+p_3} F(\hat\rho_3) \right] \\
  = p_1 F(\hat\rho_1) + (p_2+p_3)\,F\left( \frac{p_2}{p_2+p_3} \, \hat\rho_2 + \frac{p_3}{p_2+p_3} \, \hat\rho_3 \right) \\
  = F\left[ p_1 \hat\rho_1 + (p_2+p_3) \left( \frac{p_2}{p_2+p_3} \, \hat\rho_2 + \frac{p_3}{p_2+p_3} \, \hat\rho_3 \right) \right] \\
  = F\left( p_1 \hat\rho_1 + p_2 \hat\rho_2 + p_3 \hat\rho_3 \right) 
\end{multline*}
(here we kept in mind that $p_1+p_2+p_3=1$).

Finally, let us join together Eqs.~(\ref{eq:total-probability-N2}), (\ref{eq:main-N}) and (\ref{eq:rho-general}):
\begin{equation} \label{eq:total-probability-N3}
  {\mathbb E} = F \left( \sum_{k=1}^N p_k \hat\rho_k \right) \equiv F(\hat\rho) .
\end{equation}
Therefore Eq.~(\ref{eq:F-def}) is valid not only for pure states and improper mixtures (that was established in Section~\ref{improper1}), 
but also for \emph{any} states. That is, in any case the expected value ${\mathbb E}$ depends on the state of the measured system only through its 
density matrix $\hat\rho$.

Though we considered only measurements by some definite measuring apparatus~${\cal A}$, no special requirements were imposed on~${\cal A}$. 
So our reasoning does not depend on the kind of the measuring apparatus. Thus we conclude that, 
if two states $s_1$ and $s_2$ of some quantum system ${\cal S}$ are described by the same density matrix, 
then expected values of all possible measurements on ${\cal S}$ in state $s_1$ must be the same as in state $s_2$. 
(Of course, we consider only measurements on system ${\cal S}$ \emph{alone}.)

The probability of occurring a particular measurement result can be considered as a specific case of expected value. 
Indeed, let $X$ denote an outcome of some measurement (a random variable), and $\alpha$ be some fixed quantity. 
Let us define the variable $x_\alpha$ that is equal to 1 if $X=\alpha$, and equal to 0 otherwise. 
Obviously, if the apparatus measures $X$ then it also measures $x_\alpha$, and the expected value of $x_\alpha$ is the same as 
the probability that $X$ takes the value $\alpha$. Therefore any conclusions about expected values can also be applied to probabilities. 
Hence, for any two states of system~${\cal S}$ having the same density matrix,
probabilities of all possible results of whichever measurement on ${\cal S}$ also must be the same.
In other words, if two states have the same density matrix, then all statistical properties of these states are the same, 
i.~e. there is no way to distinguish experimentally between these states. 

Thus, Statement~$\rho$ is proven, and thereby the objective of this paper is achieved.


\section{Discussion}\label{discussion}

First of all, it follows from Statement~$\rho$ that any measurement is described by a POVM~\cite{Holevo2013}. 
Indeed, let us consider a mixture $\mathfrak{m}$ of two states $s_1$ and $s_2$ of some quantum system, which are taken with probabilities $p$ 
and $1-p$ respectively. If some experimental yes/no test of the quantum system is passed with probabilities $P(s_1)$ and $P(s_2)$ in 
states $s_1$ and $s_2$ respectively, then the probability $P(\mathfrak{m})$ of passing the test in the mixture is, according to 
the law of total probability,
\begin{equation} 
  P(\mathfrak{m}) = p \, P(s_1) + (1-p) \, P(s_2).
\end{equation}
On the other hand, density matrices of the states $\hat\rho(s_1)$, $\hat\rho(s_2)$ contribute into the 
density matrix of the mixture $\hat\rho(\mathfrak{m})$ in the same proportions $p$ and $1-p$:
\begin{equation} 
  \hat\rho(\mathfrak{m}) = p \, \hat\rho(s_1) + (1-p) \, \hat\rho(s_2) .
\end{equation}
Hence the probability $P$ of passing the test is an \emph{affine} function of the density matrix. 
Then, due to Theorem~2.6 of Ref.~\onlinecite{Holevo2013}, there exists a positive Hermitian operator $\hat M$ such that, 
for every density matrix $\hat\rho$,
\begin{equation} 
  P(\hat\rho) = \mathrm{Tr} (\hat\rho \hat M) .
\end{equation}
Occurring of a definite outcome of a measurement can be considered as a test. 
Therefore for a given measurement apparatus there is a collection of positive operators that correspond to different outcomes. 
This collection is a POVM. 
Hence our results must convince anyone who accepts Assumptions $\psi$, $U$, $I$, $P$, $M$ formulated above, that a quantum measurement 
of the most general type is described by a POVM.

Our study shows that Statement~$\rho$ can be justified without any use of Born rule. 
This opens the possibility to \emph{derive} Born rule from other postulates of quantum mechanics (not related to measurement) 
and from Assumptions $\psi$, $U$, $I$, $P$, $M$. Let us illustrate this on the example of measuring of the electron spin, 
whose state can be described by the polarization vector $\mathbf{p}$ defined according to Eq.~(\ref{eq:06a}). 
We consider the experiment of Stern--Gerlach type, where the electron beam divides into two branches by application of some external field; 
and there are such two states $\mathbf{p}_1$ and $\mathbf{p}_2$ that electrons with spin polarization $\mathbf{p}_1$ ($\mathbf{p}_2$) 
always take the first (the second) branch. The question is: what is the probability $P(\mathbf{p})$ of taking the first branch for the 
electron with arbitrary spin polarization $\mathbf{p}$ within the Bloch ball $|\mathbf{p}|\leq1$? 
For answering this question, it is enough to notice that (i)~$P(\mathbf{p})$ is an affine function (as discussed above) and therefore 
has the form of Eq.~(\ref{eq:07}); (ii)~the maximal and minimal values of $P$ over the Bloch ball are achieved at $\mathbf{p}_1$ 
and~$\mathbf{p}_2$ correspondingly: $P(\mathbf{p_1})=1$, $P(\mathbf{p_2})=0$. It directly follows from these two notices that 
$|\mathbf{p_1}|=|\mathbf{p_2}|=1$ (i.~e. these states are pure); moreover, $\mathbf{p_2}=-\mathbf{p_1}$ 
(i.~e. these states are orthogonal to each other), and finally, $P(\mathbf{p})=(\mathbf{p}\cdot\mathbf{p}_1+1)/2$. In the latter 
equality one can switch from polarization vectors $\mathbf{p}$, $\mathbf{p}_1$ to density matrices $\hat\rho$, $\hat\rho_1$ using Eq.~(\ref{eq:06a}), and get 
$P(\hat\rho)=\mathrm{Tr}(\hat\rho \hat\rho_1)$, i.~e. 
\begin{equation} \label{eq:born-rule-spin}
  P(\hat\rho) = \left< \psi_1 \right| \hat\rho \left| \psi_1 \right> ,
\end{equation}
where we made use of purity of the state $\hat\rho_1$ and represent this density matrix through a corresponding unit vector 
$\left| \psi_1 \right>$: $\hat\rho_1=\left| \psi_1 \right> \left< \psi_1 \right|$. 
So we have derived Born rule for measuring the electron spin, Eq.~(\ref{eq:born-rule-spin}). 
It is possible to generalize this way of justifying Born rule to the case of arbitrary finite-dimensional Hilbert space---see 
Appendix~F in Ref.~\onlinecite{nenashev2014env2}.

Note that, as a by-product, we have established purity and mutual orthogonality 
of states $\mathbf{p}_1$ and $\mathbf{p}_2$, which correspond to different outcomes of the measurement. 
Therefore one need not to \emph{postulate} that eigenvalues of any observable form an orthonormal set of vectors, i.~e. that observables correspond to Hermitian operators. The question ``why observables are 
described by Hermitian operators'' thus can be answered: ``because of Statement~$\rho$ and 
such basic property of an observable as existence of a set of states, 
for which the measurement result is predictable''.

Finally, let us briefly review other ways of introducing density matrices, and compare them 
with our approach.

(i) Statement~$\rho$ can be simply postulated~\cite{Kraus,Holevo}. This provides the shortest way to such concepts as 
POVMs, Kraus maps, etc., but gives no answer to the question whether the statistical interpretation of quantum mechanics 
is unique. 
A similar approach, in the framework of the axiomatic approach to quantum theory, consists in  
postulating that the quantum system's state is defined by only finite number of parameters, and then identifying this set of 
parameters with the density matrix~\cite{Hardy2001,Hardy2013}.

(ii) The usual ``textbook'' way consists in constructing the density matrices by mixing pure states~\cite{Dirac}, 
or by considering a state of a part of a composite system~\cite{Landau}. 
In this approach, Statement~$\rho$ is derived from the Born rule, and only for measurements of observables. 
Applicability of Statement~$\rho$ to all possible measurements stays beyond the consideration.

(iii) Statement~$\rho$ can be derived from possibility of reducing different proper mixtures to the same improper one. 
As it follows from a lemma due to Gisin~\cite{Gisin1989}, if two proper mixtures have the same density matrix, 
they can be prepared from the same pure state of a composite system by measuring a remote subsystem---see
discussion in Ref.~\onlinecite{Weinberg2014}. However, this method is based on presumption that Born rule is valid, 
therefore it does not provide an opportunity to prove Born rule. 

(iv) Von Neumann in his early paper~\cite{Neumann1927} proposed a deductive reconstruction of the probabilistic 
structure of quantum theory (see also chapter IV in book~\cite{Neumann}). 
From very general consideration, he had obtained Eq.~(\ref{eq:born-rule-rho}) for measurement of observables 
(and therefore Born rule as a particular case). Von Neumann's derivation is based on rather strong assumptions 
about observables. For example, he postulated that the expected value of the sum $\hat x + \hat y$ is always 
equal to the sum of expected values of $\hat x$ and $\hat y$, even if observables $\hat x$ and $\hat y$ are incompatible. 
Our approach can be considered as dual to von Neumann's one: we do not put any constraints on measurable quantities, 
but make use of some properties of states, that follow from the thought experiments shown in Fig.~\ref{fig:3}.

(v) The famous Gleason's theorem~\cite{Gleason1957} also enables one to get Eq.~(\ref{eq:born-rule-rho}) 
and thus to introduce the concept of density matrix. But this approach is restricted to measurements of observables, 
and requires non-contextuality of measurements (that is hard to justify). Notice also a possibility of a ``reverse'' 
use of Gleason's theorem~\cite{Nenashev2014}, in which the experimental test is fixed and states of the quantum system are varied 
(unlike the usual setting where the state is fixed and tests are varied).

We also emphasize that, like Zurek's works on \emph{envariance}~\cite{Zurek2003,Zurek2005}, our study discloses the significant role of entanglement in 
establishing the rules for probabilities in quantum mechanics.


\section{Conclusions}\label{conclusions}

This paper is devoted to Statement~$\rho$, which claims that all information about the state of a 
quantum system (i.~e. information about probabilities of results of all possible 
measurements) is encoded in the system's density matrix. 
We have shown that Statement~$\rho$ can be proven without using any probabilistic postulates, like 
Born rule. Our proof is based on thought experiments shown in Fig.~\ref{fig:3}, completed by  
Zurek's \emph{envariance} argument~\cite{Zurek2003,Zurek2005}. 
The derivation of Statement~$\rho$ depends only on quite reasonable Assumptions $\psi$, $U$, $I$, $P$, $M$, 
and on the non-measurement part of quantum mechanics. This provides also a possibility to justify the Born rule, 
and to explain the connection between observables and Hermitian operators.


\appendix
\section{Proof of Eq.~(\ref{eq:main}) without the continuity assumption}\label{app}

In this Appendix we will suppose that
\begin{equation} \label{eq:a-f-choice}
  F(\hat\rho_0) \leq F(\hat\rho_1) \, .
\end{equation}
This does not lead to any loss of generality, because $\hat\rho_0$ and $\hat\rho_1$ enter into Eq.~(\ref{eq:main}) symmetrically. 

Let us define a set $\Lambda\subset[0,1]$ of rational numbers of the form $p/2^q$, where $q=1,2,3,\ldots$ and $p=0,1,\ldots,2^q$. 
In other words, $\Lambda$ is the set of numbers between 0 and~1 whose representation as a binary fraction has a finite length. 
It was shown in Section~\ref{improper2} on the basis of Eq.~(\ref{eq:f-one-half}) that the equality~(\ref{eq:main}) 
is fulfilled for any $\lambda\in\Lambda$. Therefore
\begin{equation} \label{eq:a-rho-x-property}
  \forall x\in\Lambda, \quad F(\hat\rho_x) = (1-x) F(\hat\rho_0) + x F(\hat\rho_1) \, ,
\end{equation}
where we introduce a shorthand notation
\begin{equation} \label{eq:a-rho-x-def}
  \hat\rho_x  \stackrel{\text{def}}{=}  (1-x) \hat\rho_0 + x \hat\rho_1 .
\end{equation}
Now let us choose some real number $\lambda$ such as
\begin{equation} \label{eq:a-lambda-choice}
  0<\lambda<1 ,
\end{equation}
and define a subset $\Lambda_{-}$ of numbers in $\Lambda$, which are less then $\lambda$,
\begin{equation} \label{eq:a-lambda-minus-def}
  \Lambda_{-} = \{ x\in\Lambda :  x<\lambda \} ,
\end{equation}
and another subset $\Lambda_{+}$ of numbers in $\Lambda$, which are greater then $\lambda$,
\begin{equation} \label{eq:a-lambda-plus-def}
  \Lambda_{+} = \{ x\in\Lambda :  x>\lambda \} .
\end{equation}
Then, we choose arbitrarily two numbers $\xi\in\Lambda_{-}$ and $\eta\in\Lambda_{+}$. By definition of $\Lambda_{-}$ and $\Lambda_{+}$, 
\begin{equation} \label{eq:a-xi-lambda-eta}
  \xi < \lambda < \eta 
\end{equation}
and, due to Eqs.~(\ref{eq:a-f-choice}) and (\ref{eq:a-rho-x-property}),
\begin{equation} \label{eq:a f-xi-f-eta}
  F(\hat\rho_\xi) \leq F(\hat\rho_\eta) \, . 
\end{equation}

Now we return to the thought experiments shown in Fig.~\ref{fig:3} and modify them in the following way. 
Let us select such two bipartite states~$|\widetilde\Psi_0\rangle$ and~$|\widetilde\Psi_1\rangle$ 
of the combined system~$A+B$, that the~reduced density matrix of~$A$ 
is equal to~$\hat\rho_\xi$ for the state~$|\widetilde\Psi_0\rangle$, and to~$\hat\rho_\eta$ for the state~$|\widetilde\Psi_1\rangle$. 
We modify source $S$ so that it will emit a pair of particles $A,B$ in the joint state $|\widetilde\Psi_0\rangle$ 
instead of $|\Psi_0\rangle$. Also we modify gate $G$ so that it will operate according to rules similar to 
Eqs.~(\ref{eq:g0}),~(\ref{eq:g1}), but with $|\widetilde\Psi_0\rangle$ and~$|\widetilde\Psi_1\rangle$ 
instead of $|\Psi_0\rangle$ and~$|\Psi_1\rangle$. 
Finally, source $Q$ in the modified experiments will emit a pair of qubits $\alpha,\beta$ in the state
\begin{equation} \label{eq:a-phi}
  |\widetilde\Phi\rangle =  \sqrt{\frac{\eta-\lambda}{\eta-\xi}} \; |0\rangle |0\rangle +  \sqrt{\frac{\lambda-\xi}{\eta-\xi}} \; |1\rangle |1\rangle .
\end{equation}
The rest of the setup will be left unchanged. 

Let us analyse the expectation value of the particle~$A$'s measurement result, just like it was done 
Section~\ref{improper2}, but for the modified setup. It is easy to see that the 
reduced density matrix of particle~$A$ after passing through the gate in the first experiment is 
\begin{equation} \label{eq:a-rho-lambda-via-xi-eta}
  \frac{\eta-\lambda}{\eta-\xi} \, \hat\rho_\xi  +  \frac{\lambda-\xi}{\eta-\xi} \, \hat\rho_\eta .
\end{equation}
According to definition~(\ref{eq:a-rho-x-def}), this value is equal to $\hat\rho_\lambda$. Therefore the 
expectation value of measurement result is equal to $F(\hat\rho_\lambda)$. 
Then, almost literally repeating the reasoning between Eq.~(\ref{eq:p-m-flash-a}) and Eq.~(\ref{eq:f-p1}), 
one can obtain a relation analogous to Eq.~(\ref{eq:f-p1}):
\begin{equation} \label{eq:a-f-lambda-xi-eta-1}
  F(\hat\rho_\lambda) = (1-\tilde a) \, F(\hat\rho_\xi) + \tilde a \, F(\hat\rho_\eta) \, ,
\end{equation}
where $\tilde a$ is the probability of obtaining 1 by meter $\mu$ in the modified setup. 
As $\tilde a \in [0,1]$, the latter equation means that the quantity $F(\hat\rho_\lambda)$ is between 
$F(\hat\rho_\xi)$ and $F(\hat\rho_\eta)$. Therefore, taking Eq.~(\ref{eq:a f-xi-f-eta}) into account, 
one can conclude that 
\begin{equation} \label{eq:a-f-lambda-xi-eta-2}
   F(\hat\rho_\xi) \leq F(\hat\rho_\lambda) \leq F(\hat\rho_\eta) .
\end{equation}

Since these inequalities were derived for arbitrarily chosen $\xi\in\Lambda_{-}$ and $\eta\in\Lambda_{+}$, 
then for \emph{any} $\xi\in\Lambda_{-}$ the value of $F(\hat\rho_\xi)$ is less than $F(\hat\rho_\lambda)$. 
That is, $F(\hat\rho_\lambda)$ is an upper bound of the set $\{ F(\hat\rho_\xi) : \xi\in\Lambda_{-} \}$. Therefore, 
\begin{equation} \label{eq:a-f-lambda-sup}
   F(\hat\rho_\lambda)  \geq  \sup \, \{ F(\hat\rho_\xi) : \xi\in\Lambda_{-} \} \, . 
\end{equation}
Similarly,
\begin{equation} \label{eq:a-f-lambda-inf}
   F(\hat\rho_\lambda)  \leq  \inf \, \{ F(\hat\rho_\eta) : \eta\in\Lambda_{+} \} \, .
\end{equation}

Right-hand sides of the latter two inequalities can be easily calculated. 
Since $\Lambda$ is a dense subset of the range $[0,1]$, then
\begin{equation} \label{eq:a-sup-inf-lambda}
  \sup \Lambda_{-}  =  \inf  \Lambda_{+}  =  \lambda \, .
\end{equation}
It follows from Eqs.~(\ref{eq:a-f-choice}), (\ref{eq:a-rho-x-property}) and (\ref{eq:a-sup-inf-lambda}) 
that if $x\in\Lambda_{-}$ then $F(\hat\rho_x)$ is not greater than the quantity $(1-\lambda) F(\hat\rho_0) + \lambda F(\hat\rho_1)$, 
but can be made arbitrarily close to this quantity by appropriate choice of $x$. Hence,
\begin{equation} \label{eq:a-sup-f-lambda-minus}
  \sup \, \{ F(\hat\rho_x) : x\in\Lambda_{-} \}  =  (1-\lambda) F(\hat\rho_0) + \lambda F(\hat\rho_1) \, .
\end{equation}
In a similar way,
\begin{equation} \label{eq:a-inf-f-lambda-plus}
  \inf \, \{ F(\hat\rho_x) : x\in\Lambda_{+} \}  =  (1-\lambda) F(\hat\rho_0) + \lambda F(\hat\rho_1) \, .
\end{equation}
Finally, one can easily find $F(\hat\rho_\lambda)$ from Eqs.~(\ref{eq:a-f-lambda-sup}), (\ref{eq:a-f-lambda-inf}), 
(\ref{eq:a-sup-f-lambda-minus}) and (\ref{eq:a-inf-f-lambda-plus}):
\begin{equation} \label{eq:a-f-lambda-answer}
   F(\hat\rho_\lambda)  =  (1-\lambda) F(\hat\rho_0) + \lambda F(\hat\rho_1) \, .
\end{equation}
Remembering the definition of $\hat\rho_\lambda$, Eq.~(\ref{eq:a-rho-x-def}), one can see that Eq.~(\ref{eq:a-f-lambda-answer}) 
is exactly the same as Eq.~(\ref{eq:main}). Thus, Eq.~(\ref{eq:main}) is proved for any $\lambda\in(0,1)$.


\bibliography{envariance}

\end{document}